\documentclass[10pt]{paper}
\usepackage{amssymb,latexsym, amsmath}
\usepackage{amscd}

\newtheorem{thm}{Theorem}
\newtheorem{prop}[thm]{Proposition}
\newtheorem{lem}[thm]{Lemma}

\newtheorem{remark}{Remark}
\newtheorem{exm}{Example}

\newcommand{\C}{\mathbb{ C}}
\newcommand{\R}{\mathbb{ R}}

\newcommand{\g}{\mathfrak{ g}}

\DeclareMathOperator{\pr}{pr}

\DeclareMathOperator{\tr}{tr}
\DeclareMathOperator{\diag}{diag}

\DeclareMathOperator{\Ad}{Ad}
\DeclareMathOperator{\rank}{rank}
\DeclareMathOperator{\ad}{ad}

\begin{document}
\title{Three Natural Mechanical  Systems on Stiefel Varieties}

\author{Yuri N. Fedorov \and Bo\v zidar Jovanovi\' c}

\maketitle

\begin{abstract}
We consider integrable generalizations of the spherical pendulum
system to the Stiefel variety $V(n,r)=SO(n)/SO(n-r)$ for a certain metric.
For the case of $V(n,2)$ an alternative integrable model
of the pendulum is presented.

We also describe a system on the Stiefel variety with a four-degree potential. The latter has invariant relations on
$T^*V(n,r)$ which provide the complete integrability of the flow
reduced on the oriented Grassmannian variety $G^+(n,r)=SO(n)/SO(r)\times SO(n-r)$.
\end{abstract}

\section{Introduction}

The {\it Stiefel variety} $V(n,r)=SO(n)/SO(n-r)$ is the variety of
ordered sets of $r$ orthogonal unit  vectors $e_1,\dots, e_r$ in
the Euclidean space $\R^n$, or, equivalently, the set of $n \times
r$ matrices
$$
X=(e_1 \cdots e_r) \in M_{n,r}(\R)
$$
satisfying ${X}^T {X}={\bf I}_r$, where ${\bf I}_r$ is the
$r\times r$ unit matrix. In this paper we follow notations used in
\cite{FeJo3}, where the integrability and geometrical properties
of geodesic and Neumann flows on Stiefel varieties have been
studied (Section 2).

The spherical pendulum is one of the simplest integrable systems.
It can be defined as a natural mechanical system with the Hamiltonian
\begin{equation}
H =\frac12\langle p,p\rangle +\langle \gamma,x\rangle,
\label{spherical}
\end{equation}
where $\gamma$ is a fixed vector in $\R^n$ and the cotangent
bundle $T^*S^{n-1}$ is realized as a submanifold of
$\R^{2n}\{x,p\}$ given by constraints $\langle x,x\rangle =1$,
$\langle x,p\rangle=0$.

By the analogy, define the {\it pendulum} on the Stiefel variety
$V(n,r)$ as a natural mechanical system with a $SO(n)\times
SO(r)$-invariant kinetic energy and a linear potential
\begin{equation}
V_{pendulum}(X)=\tr(X^T\Gamma)=\sum_{i=1}^r \langle
e_i,\gamma_i\rangle, \label{pendulum_potential}
\end{equation}
where $\Gamma=(\gamma_1\cdots\gamma_r)$ is a constant $n\times r$
matrix  (Section 3). For a certain $SO(n)\times SO(r)$-invariant
metric, we present a Lax representation of the system and show that
the equations can be regarded as a subsystem of an integrable
Hamiltonian system on the dual of the semi-direct product of
$so(n)\oplus so(r)$ and of the vector space $M_{n,r}(\R)$, which
was considered by Reyman \cite{R} and Bolsinov \cite{Bo}.

We also give the Lax representation in the case when all the vectors $\gamma_i$ are
collinear and the kinetic energy is determined by a normal metric on $V(n,r)$.

For $r=2$ there is another natural choice of the pendulum-type potential
\begin{equation}\label{pmp}
\tilde V_{pendulum}(e_1,e_2)=-\langle e_1 \wedge e_2, \Xi
\rangle=-\tr(e_1e_2^T\Xi),
\end{equation}
where $\Xi$ is a fixed element of $so(n)$ (Section 4). We prove
its complete integrability using the results of \cite{BJ4}.

Finally, we consider a system on $V(n,r)$ with a four-degree
potential, which for $r=1$ is described by the Hamiltonian
\begin{equation}
H=\frac12\langle p,p\rangle+\nu\left(\sum_{i=1}^n a_i^2
x_i^2-\left(\sum_{i=1}^n a_i x_i^2 \right)^2\right),\label{4sp}
\end{equation}
$\nu$ being an arbitrary parameter. The system on $S^{n-1}$ is separable in
elliptical (spheroconical) coordinates (see Wojciechowski \cite{Wo}).\footnote{The
separability is also preserved when we add the Neumann potential
$\sum_i a_i x_i^2$. We do not treat the generalization of this
system here.} Later, in \cite{S1, S2} Saksida considered {\it
generalized Neumann systems} on Cartan models of symmetric spaces,
and in the particular case of the projective space
$\mathbb{RP}^{n-1}$ he recovered the Hamiltonian \eqref{4sp}.

We prove that $V(n,r)$-analog of the above system, when reduced to
the oriented Grassmannian variety $G^+(n,r)$, is also completely
integrable (Section 5). In our proof we used the mentioned
construction of \cite{S1, S2}.

\section{Hamiltonian flows on Stiefel varieties}
The tangent bundle $TV(n,r)$ is canonically defined by
\begin{equation}
\label{cond_XX} {X}^T {X} ={\bf I}_r, \quad {X}^T {\dot X} + {\dot
X}^T {X} =0\, ,
\end{equation}
while the cotangent bundle $T^*V(n,r)$ can be realized as the set
of pairs of $n\times r$ matrices $({X},{P})$ satisfying the
constraints
\begin{equation}
\label{cond_XP} {X}^T {X} ={\bf I}_r, \quad {X}^T {P} + {P}^T {X}
=0\, ,
\end{equation}
which give $r(r+1)$ independent scalar constraints. The canonical
symplectic structure $\omega$ on $T^*V(n,r)$ is the restriction of
the canonical 2-form from the ambient space $T^*M_{n,r}(\R)$. It
is convenient to work in redundant variables $(X,P)$ when the
canonical Poisson structure $\{\cdot,\cdot\}$ on $T^*V(n,r)$ is
defined by the use of the Dirac construction \cite{Dirac, Moser}.
Namely, let $\{\cdot,\cdot\}_{T^*M_{n,r}(\R)}$ be the canonical
Poisson bracket on $\R^{2nr}$
$$
\{f_1,f_2\}_{T^*M_{n,r}(\R)}=\sum_{i=1}^r \left( \left\langle
\frac{\partial f_1}{\partial e_i}, \frac{\partial f_2}{\partial
p_i} \right\rangle -\left\langle \frac{\partial f_1}{\partial
p_i},\frac{\partial f_2}{\partial e_i}\right\rangle\right)
$$
and $C_{i,j}$ be the inverse of the matrix
$\{F_i,F_j\}_{T^*M_{n,r}(\R)}$, $i,j=1,\dots,r(r+1)$, where we
denoted constraints (\ref{cond_XP}) by $F_i=0$,
$i=1,\dots,r(r+1)$. Then the Dirac bracket is given by
\begin{equation*}
\{f_1,f_2\}=\{f_1,f_2\}_{T^*M_{n,r}(\R)} +\sum_{i,j}
\{F_i,f_1\}_{T^*M_{n,r}(\R)} C_{i,j} \{F_j,f_2\}_{T^*M_{n,r}(\R)}.
\label{Dirac_bracket}
\end{equation*}
The subvariety $T^*V_{n,r}$ appears as a symplectic leaf of the
Dirac bracket and the restriction of $\{f_1,f_2\}$ to $T^*V_{n,r}$
depends only on the restriction of $f_1$ and $f_2$ to
$T^*V_{n,r}$. A slightly different realization of the cotangent
bundle $T^*V(n,r)$ is given in \cite{BCS}.

With the above notation, the Hamiltonian equation $ \dot f=\{f,H\}
$ in the matrix form reads
\begin{equation}
\begin{aligned}
&\dot X=\frac{\partial H}{\partial P} - X\Pi, \\
&\dot P=- \frac{\partial H}{\partial X} + X\Lambda + P\Pi
\end{aligned}\label{dot_XP}
\end{equation}
The $r\times r$ symmetric Lagrange multipliers matrices $\Lambda$
and $\Pi$ are uniquely determined from the condition that the
trajectory $(X(t),P(t))$ satisfies the constraints
(\ref{cond_XP}).

The Lie groups $SO(n)$ and $SO(r)$ naturally act on $T^*V(n,r)$ by
left and right multiplications respectively, with the equivariant
momentum mappings given by (e.g., see \cite{FeJo3})
\begin{eqnarray}
&&\Phi: T^*V(n,r) \to so(n)^*, \qquad \Phi=P X^T-X P^T\,(=p_1\wedge e_1+\dots+p_r\wedge e_r), \label{momentum_map} \\
&&\Psi: T^*V(n,r) \to so(r)^*, \qquad \Psi=X^T P - P^T X.
\label{momentum_right_map}
\end{eqnarray}
Here we identified $so(n)\cong so(n)^*$ and  $so(r)\cong so(r)^*$
by the use of the invariant metrics on $so(n)$ and $so(r)$ defined
by $\langle \eta_1,\eta_2\rangle=-\frac12 \tr(\eta_1 \eta_2)$.

Consider a family of natural mechanical systems with the
$SO(n)\times SO(r)$-invariant kinetic energy
\begin{eqnarray*}
T_\kappa(X,P)&=& \frac12\langle \Phi,\Phi\rangle +\frac12 \kappa
\langle\Psi,\Psi\rangle \\
 &=& \frac12\tr(P^TP)-\left(\frac12+\kappa\right)\tr((X^T P)^2),
 \end{eqnarray*}
$\kappa$ being a parameter ($\kappa>-1$).  In the family of the
metrics $ds^2_\kappa$ determined by $T_\kappa$
there is the {\it normal metric} induced from a bi-invariant metric
on $SO(n)$ ($\kappa=0$) and the {\it Euclidean metric} induced
from the Euclidean metric of the ambient space $M_{n,r}(\R)$
($\kappa=-1/2$). We refer to generic $ds^2_\kappa$ as {\it Jensen
metrics}. For $r=2$ there is a unique $\kappa$, while for $r>2$
there are exactly two values of $\kappa$, such that $ds^2_\kappa$
is an Einstein metric (see \cite{Je}).

We note that Hamiltonian equations considered in the paper are
equivalent to the second-order Euler--Lagrange equations with
multipliers via the Legendre transformation.

\begin{lem}
The Legendre and the inverse Legendre transformation determined by
the Jensen metric $ds^2_\kappa$ are given by
\begin{equation}\label{legendre}
P=\dot X-\frac{1+2\kappa}{2+2\kappa} XX^T\dot X,\qquad \dot
X=P-(1+2\kappa) X P^TX,
\end{equation}
respectively. Here $(X,\dot X)$ and $(X,P)$ are subject to the
constraints \eqref{cond_XX} and \eqref{cond_XP}.
\end{lem}

\begin{remark}
For $\kappa=-1$, the kinetic energy $T_{-1}(X,P)$ defines the
sub-Riemannian metric $ds^2_{-1}$ on $SO(n)\times SO(r)$-invariant
nonintegrable distribution $D\subset TV(n,r)$ given by the
constraints
$$
X^T\dot X=0.
$$
\end{remark}

If the potential $U(X)$ defining a natural mechanical system
$H_\kappa=T_\kappa+U$ is $SO(r)$-invariant, one can also perform a
reduction of the system to a natural mechanical system on the {\it
oriented Grassmannian} $G^+(n,r)$,  the variety of $r$-dimensional
oriented planes passing through the origin in $\R^n$. It can be
seen as a quotient of the Stiefel manifold by the right
$SO(r)$-action via submersion
\begin{equation}
\pi: V(n,r) \to G^+(n,r), \quad \pi(e_1\,e_2\dots e_r)=e_1\wedge
e_2\wedge \dots \wedge e_r. \label{submersion}
\end{equation}

The symplectic leafs of the Poisson manifolds $(T^*V(n,r))/SO(r)$
are the Marsden-Weinstein symplectic reduced spaces of $T^*
V(n,r)$. The reduced space that corresponds to the zero value of
the momentum mapping, $\Psi^{-1}(0)/SO(r)$, is symplectomorphic to
the cotangent bundle $T^* G^+(n,r)$ equipped with the canonical
symplectic structure (see e.g., \cite{GS}).  Note that the reductions of the
systems $H_\kappa=T_\kappa+U$ to $T^* G^+(n,r)$ have the same
kinetic energy given by a {\it normal metric} on $G^+(n,r)$ for
all $\kappa$.

\begin{exm} {\rm The celebrated Neumann system
on the sphere $S^{n-1}$ is defined as a natural mechanical system
with a $SO(n)$-invariant metric and a quadratic Hamiltonian.
Analogously, paper \cite{FeJo3} considered the Neumann system on $V(n,r)$ with
a Jensen metric $ds^2_\kappa$ and $SO(r)$-invariant quadratic
potential energy $H_{neumann,\kappa}=T_\kappa+\frac12\tr(X^TAX)$,
$A=\diag(a_1,\dots,a_n)$. The system is right
$SO(r)$-invariant. As shown in \cite{FeJo3}, it is completely integrable in the
non-commutative sense by means of the integrals
\begin{equation}
\{ \tr(\lambda(PX^T-XP^T)+XX^T-\lambda^2 A)^k\, \vert\,
k=1,\dots,n, \, \lambda\in\mathbb{R}\} \label{perelomov}
\end{equation}
and by the components of the momentum mapping $\Psi$. On the other hand, the reduced system on $T^*
G^+(n,r)$ is completely integrable in the Liouville sense by means
of the integrals induced from the $SO(r)$-invariant functions
(\ref{perelomov}) (see also \cite{FeJo3}).}
\end{exm}

\section{$V(n,r)$-pendulum}

Consider the pendulum system determined by the kinetic energy
which corresponds to the Jensen metric $ds^2_{\kappa}$ and the
potential (\ref{pendulum_potential}):
\begin{equation}
H_{pendulum,\kappa}(X,P)=\frac12\tr(P^TP)-\frac{(1+2\kappa)}2\tr((X
P^T)^2)+\tr(X^T\Gamma), \label{spherical_pendulum}
\end{equation}
where, as above, $\Gamma=(\gamma_1\cdots\gamma_r)$ is a constant $n\times r$
matrix.

For $r=1$, the Hamiltonian (\ref{spherical_pendulum}) coincides
with (\ref{spherical}), while for $r=n$ it represents the
Hamiltonian function for the motion of a symmetric rigid body in
the presence of $n$ homogeneous potential fields. This is the
special case of the generalized Kowalewski top considered in
\cite{RS} (one should take $q=p$ in eq. (7.29), page 168,
\cite{RS}).

Using the equations \eqref{dot_XP}, we arrive at

\begin{prop} In the redundant
coordinates $(X,P)$, the equations of the motion of the pendulum
given by the Hamiltonian function \eqref{spherical_pendulum} read
\begin{equation}
\begin{aligned}
&\dot X=P-(1+2\kappa)XP^TX, \\
&\dot P=(1+2\kappa)PX^TP-\Gamma +X\Lambda ,
\end{aligned}
\label{SP_XP}
\end{equation}
where $\Lambda=\frac12(-2P^TP+X^T\Gamma+\Gamma^T X)$ is
the matrix Lagrange multiplier.
\end{prop}

Since the metrics $ds^2_\kappa$ are $SO(n)\times SO(r)$-invariant,
the system \eqref{SP_XP} has the Noether integrals with respect to
the infinitesimal symmetries of the $SO(n)\times SO(r)$-action
that leave the potential (\ref{pendulum_potential}) invariant.

\subsection{$SO(n-1)$--invariant pendulum}

Now assume that in the pendulum system \eqref{spherical_pendulum}
$$
\gamma_1=a_1 \alpha, \quad \gamma_2=a_2\alpha_2,\quad\dots,\quad
\gamma_r=a_r\alpha, \qquad a_1^2+a_2^2+\dots+a_r^2\ne 0,
$$
where $\alpha$ is a unit vector. The potential
(\ref{pendulum_potential}) takes the form
\begin{equation}\label{pendulum-alpha}
V_{pendulum}(e_1,\dots,e_r)=\langle a_1 e_1+\dots+a_r
e_r,\alpha\rangle.
\end{equation}

Let $SO(n)_\alpha\cong SO(n-1)$  and $so(n)_\alpha\cong so(n-1)$
be the the isotropy group and the isotropy algebra of $\alpha$,
respectively. Then the potential  (\ref{pendulum-alpha}) is
$SO(n)_\alpha$-invariant. The resulting Noether's integrals are
the components of the momentum mapping
\begin{equation}\label{phi-a}
\Phi_\alpha=\Phi(X,P)-\pr_{\alpha\wedge\R^n}\Phi(X,P)=\Phi-(\Phi\alpha)\wedge\alpha:\quad
T^*V(n,r)\to so(n)_\alpha^*.
\end{equation}

The dimension $s$ of the image $\Phi_\alpha(T^*V(n,r))$, i.e., the
number of independent functions within the algebra of polynomial
Noether integrals $\Phi^*_\alpha(\R[so(n)_\alpha^*])$, is the same
as the number of independent Noether integrals for
$SO(n-1)$-action on $T^*V(n-1,r)$.

Let $p_i(\xi)=\tr(\xi^{2i})$. Within the algebra
$\Phi^*_\alpha(\R[so(n)_\alpha^*])$, we have Casimir functions
$$
I_i=\Phi_\alpha\circ p_i=\tr(\Phi_\alpha^{2i}), \qquad
i=1,\dots,\left[\frac{n-1}2\right],
$$
which commute with all the Noether integrals.

Suppose $n-1\ge 2r$. Then $I_1,\dots,I_r$ are independent (see
\cite{FeJo3}). The invariants
$p_1(\xi)=\tr(\xi^2),\dots,p_r(\xi)=\tr(\xi^{2r})$ separate
generic coadjoint orbits in the image $\Phi_\alpha(T^*V(n,r))$. On
the other hand, the dimension of generic orbit in
$\Phi_\alpha(T^*V(n,r))$ is $2r(n-r-2)$ (see \cite{BJ}).
Therefore, the number of independent Noether integrals is
$$
s=2r(n-r-2)+r=2rn-2r^2-3r.
$$

Since $\dim T^*V(n,r)=2nr-r^2-r$, the generic invariant level sets
$\Phi_\alpha=const$, $H_{pendulum,\kappa}=const$ are of dimension
$r(r+2)-1$. In particular, for $r=1$ we recover the fact that the
invariant manifolds of the spherical pendulum are 2-dimensional.

We can always find
$$
s_0=(s-r)/2=r(n-r-2)
$$
polynomials $f_1,\dots,f_{s_0}$ that form a complete commutative
set on a a generic coadjoint orbit in $\Phi_\alpha(T^*V(n,r))$
with respect to the standard Lie-Poisson bracket on
$so(n)^*_\alpha$ (for example, by using the Mischenko-Fomenko
method of the argument shift, see \cite{Bo}). Therefore,
$$
I_1=\Phi_\alpha\circ p_1, \quad \dots, \quad I_r=\Phi_\alpha\circ
p_r,\quad F_1=\Phi_\alpha\circ f_1,\quad\dots, \quad
F_{s_0}=\Phi_\alpha\circ f_{s_0}
$$
are independent commuting functions on $T^*V(n,r)$. Thus, apart of
the Hamiltonian and the Noether integrals, for the complete
integrability of the $V(n,r)$-pendulum with the potential \eqref{pendulum-alpha}
we need $\frac12r(r+1)-1$ additional commuting independent integrals.

\subsection{The Lax representations}
Now we turn back to $V(n,r)$-pendulum with the general Hamiltonian
\eqref{spherical_pendulum}.
\begin{lem}
In view of the momentum mappings \eqref{momentum_map} and
\eqref{momentum_right_map}, the equations \eqref{SP_XP} are
equivalent to the following system
\begin{eqnarray}
&&\dot X=\Phi X+\kappa X\Psi, \nonumber\\
&&\dot \Phi=X \Gamma^T-\Gamma X^T \,(=e_1\wedge \gamma_1+\dots+e_r \wedge \gamma_r),\label{SPk}\\
&&\dot \Psi=\Gamma^T X- X^T \Gamma\nonumber.
\end{eqnarray}
\end{lem}

The symmetry of the above system leads to the following theorem.

\begin{thm} \label{LA_pair} If $\kappa=1$ or  $\Gamma=0$
(the case of the geodesic flow of the metric $ds^2_\kappa$),
the equations \eqref{SP_XP} are equivalent to the matrix equations
\begin{equation}\label{LASP}
\frac{d}{dt} L(\lambda)=[L(\lambda),A(\lambda)]
\end{equation}
with a spectral parameter $\lambda$, where $L(\lambda)$ and
$A(\lambda)$ are $so(n+r)$ matrices given by:
$$
L(\lambda)=\begin{pmatrix}
-\lambda\Phi & X+\lambda^2\Gamma \\
-X^T-\lambda^2\Gamma^T & \lambda \kappa\Psi
\end{pmatrix}, \quad
A(\lambda)=\begin{pmatrix}
-\Phi & \lambda\Gamma \\
-\lambda\Gamma^T & \kappa\Psi
\end{pmatrix}. $$
\end{thm}

\noindent{\it Proof.} By 
comparing terms with the same degree in $\lambda$, from
\eqref{LASP} we get the equations
\begin{eqnarray*}
&&\dot X=\Phi X+\kappa X\Psi, \qquad \quad \dot\Gamma=0, \\
&&\dot \Phi=X \Gamma^T-\Gamma X^T, \qquad \kappa\dot\Psi=\Gamma^T
X- X^T \Gamma,
\end{eqnarray*}
which correspond to \eqref{SPk} for $\kappa=1$ and arbitrary
$\Gamma$, or for $\Gamma=0$ and arbitrary $\kappa$.\hfill $\Box$

\

In the special case, when the potential is given by
\eqref{pendulum-alpha} and $\kappa=0$, from Lemma 3 we get the
equations
\begin{eqnarray*}
&& \dot x=\Phi x, \qquad\qquad x=a_1 e_1+\dots+ a_r e_r,\\
&& \dot\Phi= x \wedge \alpha.
\end{eqnarray*}

\begin{thm} Assume
that the potential is given by \eqref{pendulum-alpha} and
$\kappa=0$. Then the equations \eqref{SP_XP} imply the matrix
equations
\begin{equation}\label{LASP2}
\frac{d}{dt} {\tilde L}(\lambda)=[\tilde L(\lambda),\tilde
A(\lambda)], \qquad \lambda \in {\R},
\end{equation}
where $\tilde L(\lambda)$ and $\tilde A(\lambda)$ are $so(n+1)$ matrices given by:
$$
\tilde L(\lambda)=\begin{pmatrix}
-\lambda\Phi & x+\lambda^2\alpha \\
-x^T-\lambda^2\alpha^T & 0
\end{pmatrix}, \quad
\tilde A(\lambda)=\begin{pmatrix}
-\Phi & \lambda\alpha \\
-\lambda\alpha^T & 0
\end{pmatrix}. $$
\end{thm}

Therefore, the coefficients of the invariant polynomials
\begin{eqnarray}
&& f_{i,\lambda}(X,P)=\tr(L(\lambda))^{2i}, \quad
 i=1,2,\dots,\left[\frac{n+r}2\right] \label{A}\\
&& g_{i,\lambda}(X,P)=\tr(\tilde L(\lambda))^{2i}, \quad
 i=1,2,\dots,\left[\frac{n+1}2\right]
\label{A2}
\end{eqnarray}
are the first integrals of the system for $\kappa=1$ (or
$\Gamma=0$ and arbitrary $\kappa$) and $\kappa=0$ and the
potential given by \eqref{pendulum-alpha}, respectively.

Note that the integrability of the geodesic flows ($\Gamma=0$) of the Jensen metrics
$ds^2_\kappa$ is already proved in \cite{FeJo3}.
In the next subsection we will show that the integrals \eqref{A}
commute for $\kappa=1$ and generic $\Gamma$.

\subsection{Coadjoint representation of $T^*V(n,r)$}

Let $so(n+r)=so(n)\oplus so(r)+\mathfrak d$ be the orthogonal
decomposition of the Lie algebra $so(n+r)$ with respect to the
invariant scalar product $\langle\cdot,\cdot\rangle$. The pair
$(so(n+r),so(n)\oplus so(r))$ is symmetric: $[so(n)\oplus
so(r),\mathfrak d]\subset\mathfrak d$, $[\mathfrak d,\mathfrak
d]\subset so(n)\oplus so(r)$.

Consider the right $SO(n)\times SO(r)$ action on $T^*V(n,r)$
\begin{equation}
(R,Q)(X,P)=(R^{-1}XQ, R^{-1}PQ), \quad (R,Q)\in SO(n)\times SO(r).
\label{extended}
\end{equation}
Combining the momentum mapping of the action (\ref{extended}) with
the inclusion $V(n,r)\subset M_{n,r}(\R) \cong \mathfrak d$ we get
the mapping $ \Theta: T^*V(n,r) \to ((so(n)\oplus so(r))
\oplus_{\ad} \mathfrak d)^*\cong so(n+r)$  such
that
\begin{equation} \Theta(X,P)=
\begin{pmatrix}
-\Phi & X \\ -X^T & \Psi
\end{pmatrix}.
\label{Theta}
\end{equation}
This a Poisson mapping which realizes $T^*V(n,r)$ as a coadjoint
orbit (e.g., see eq. (29.11) on page 225, \cite{GS}, Proposition
5.6 in \cite{RS}, or Proposition 3 in \cite{BJ4}). Here we
identified the dual space of the semi-direct product $(so(n)\oplus
so(r)) \oplus_{\ad} \mathfrak d$ with $so(n+r)$ by using the
scalar product $\langle\cdot,\cdot\rangle$. Within this
identification, the Lie-Poisson bracket reads
\begin{eqnarray}
\{f,g\}_0(h+v) &=& \langle h,[\pr_{so(n)\oplus so(r)}\nabla
f,\pr_{so(n)\oplus so(r)}\nabla g]\rangle\nonumber\\
&& +\langle v,[\pr_{so(n)\oplus so(r)}\nabla f,\pr_\mathfrak
d\nabla g]+[\pr_\mathfrak d\nabla f,\pr_{so(n)\oplus so(r)}\nabla
g]\rangle,\label{semi-direct} \\
&& f,g: so(n+r)\to\R, \quad h\in
so(n)\oplus so(r), \,\, v\in \mathfrak d.\nonumber
\end{eqnarray}

On the other hand, from a general construction of compatible
Poisson structures related to a symmetric pair decomposition of
Lie algebras developed by Reyman  \cite{R} and Bolsinov \cite{Bo,
Bo2}, we get a complete set $\mathcal A+\mathcal B$ of polynomials
$\mathcal A$ and $\mathcal B$ on $((so(n)\oplus so(r))
\oplus_{\ad} \mathfrak d)^*$ described below.

Namely, let
$$
\gamma=\begin{pmatrix}
0 & \Gamma \\
-\Gamma^T & 0
\end{pmatrix}\in \mathfrak d\,.
$$
Then $\mathcal A$ is a commutative set of the polynomials defined
as the coefficients in $\lambda$ of
\begin{equation}
p_{i,\lambda}(h+v)=\tr(\lambda h+ v +\lambda^2 \gamma)^{2i}, \quad
i=1,2,\dots,\left[\frac{n+r}2\right], \quad h\in so(n)\oplus
so(r), \,\, v\in \mathfrak d \label{central2}
\end{equation}
and $\mathcal B$ is the set of linear functions on the isotropy
algebra of $\gamma$ within $so(n)\oplus so(r)$:
$$
{\mathrm{St}}(\gamma)=\{\, h\in so(n)\oplus so(r) \, \mid \,
[h,\gamma]=0\, \}.
$$
Moreover the functions from $\mathcal A$ and $\mathcal B$ mutually
commute $\{\mathcal A,\mathcal B\}_0=0$ (Theorem 1.5, \cite{Bo}
and Theorem 8, page 234 \cite{TF}, see also \cite{Bo2, Pa}).

Note that $\Theta^*p_{i,\lambda}$ are exactly the above integrals
$f_{i,\lambda}$ arising from the Lax representation of the system
for $\kappa=1$, while $\Theta^*\mathcal B$ are the Noether
integrals with respect to the infinitesimal symmetries of the
action (\ref{extended}), which leave the potential
(\ref{pendulum_potential}) invariant. Thus, the integrals
\eqref{A} commute.

Similarly, by considering the dual space of the semi-direct
product $so(n) \oplus \R^n$, one can prove commutativity of
integrals \eqref{A2}.


\subsection{The integrability problem}

The pendulum Hamiltonians (\ref{spherical_pendulum}) can be written in the
form $\Theta^* H_\kappa=H_\kappa\circ\Theta$, where $\Theta$ is given by \eqref{Theta} and
$$
H_\kappa(h+v)=\frac12\langle h,h \rangle+\frac{\kappa-1}2\langle
\pr_{so(r)} h,\pr_{so(r)} h \rangle+\langle \gamma, v \rangle.
$$
The corresponding Hamiltonian flows on $((so(n)\oplus so(r))
\oplus_{\ad} \mathfrak d)^*$  are given by
\begin{equation} \dot h=[v,\gamma], \qquad  \dot v=[v,h+(\kappa-1)\pr_{so(r)} h]. \label{Lie_algebra_equations}
\end{equation}

In particular, the flow with Hamiltonian $H_1$ is completely
integrable with a complete noncommutative set of integrals
$\mathcal A+\mathcal B$ on a generic symplectic leaf of the
Lie-Poisson bracket \eqref{semi-direct}. Thus, the mapping
\eqref{Theta} establishes the isomorphism between the
$V(n,r)$-pendulum \eqref{SP_XP} with $\kappa=1$ and a subsystem of
the integrable system \eqref{Lie_algebra_equations} laying on the
orbit $\Theta(T^*V(n,r))$.

Note that $\Theta(T^*V(n,r))$ is a {\it singular symplectic leaf}
in $((so(n)\oplus so(r)) \oplus_{\ad} \mathfrak d)^*$, on which
some of the polynomials $\mathcal A+\mathcal B$ became dependent.
To check the completeness of $\mathcal A+\mathcal B$ on
$\Theta(T^*V(n,r))$, i.e., to formally prove the complete
integrability of the pendulum system with $\kappa=1$ on
$T^*V(n,r)$, one needs an additional analysis.\footnote{One
possible approach is applying the completeness criterium of
\cite{Bo, Bo2} (see also \cite {Pa}). This involves a lot of
calculations, which we were unable to finish. We verified the
completeness criterium in the case when the projection of
$\Theta(X,P)$ to the isotropy algebra ${\mathrm{St}(\gamma)}$ is a
regular element of $\mathrm{St}(\gamma)$.} A careful analysis is
also needed for the pendulum system with the potential
\eqref{pendulum-alpha} and $\kappa=0$, since in this case the components of the
momentum mapping \eqref{phi-a} and the Lax pair integrals
\eqref{A2} do not form a complete set of functions on $T^*V(n,r)$.
They should be completed by a certain class of $SO(n)$-invariant
integrals.

Using the results of subsection 3.1, we can prove the complete
integrability of the system for $r=2$, $\kappa=1$ and an arbitrary
choice of $\gamma_1$ and $\gamma_2$, as well as for $\kappa=0$ and
$\gamma_1=a_1\alpha_1$, $\gamma_2=a_2\alpha_2$. Since the detailed
proofs are rather technical, we expect to present them elsewhere.


\section{$V(n,2)$-pendulum}

The sphere $S^2$ can be seen as a $SO(3)$-adjoint orbit. Then the
Hamiltonian \eqref{spherical} represents a natural mechanical
system with the kinetic energy given by a normal metric and the
potential being a linear function on the Lie algebra $so(3)$. Following this point of view,
one can consider integrable pendulum type systems on adjoint orbits
$\mathcal O(a)=\Ad_G(a)\subset\g$ of a compact semi-simple Lie
group $G$. Such systems, with the kinetic energy defined by a normal metric and
the potential energy given by a linear function on the Lie algebra, were
studied in \cite{BJ4}.

The oriented Grassmannian variety $G^+(n,2)$ can also be realized as a
$SO(n)$-adjoint orbit of the matrix
$$
E_1\wedge E_2=\begin{pmatrix}
0 & 1 & 0 & \dots & 0 \\
-1 & 0 & 0 & \dots & 0\\
\vdots &&&& \vdots \\
0 & 0 & 0 & \dots & 0
\end{pmatrix} .
$$
Following \cite{BJ4}, taking the pull-back of a linear
potential on the Lie algebra $so(n)$ with respect to the mapping
$\pi: V(n,2) \to G^+(n,2)\subset so(n)$
\begin{equation}\label{pr}
\pi(e_1,e_2)=e_1\wedge e_2,
\end{equation}
we get the potential \eqref{pmp}. For the case of $V(n,2)$ it is
convenient to use vector notation in which the constraints
\eqref{cond_XP} read:
\begin{equation*}
\label{cxp} \langle e_1,e_1\rangle =\langle e_2,e_2\rangle=1,
\quad \langle e_1,e_2\rangle=\langle e_1,p_1\rangle =\langle
e_2,p_2\rangle =\langle e_1,p_2\rangle +\langle e_2,p_1\rangle=0.
\end{equation*}
The $SO(2)$-momentum mapping is simply the scalar function
$$
\Psi_{12}=\langle e_1,p_2\rangle -\langle e_2,p_1\rangle
$$
and $T^*G^+(n,2)\thickapprox\Psi^{-1}_{12}(0)/SO(2)$.

The Hamiltonian equations defined by the Hamiltonian of the
pendulum systems with the metric $ds^2_\kappa$
\begin{equation}
\tilde H_{pendulum, \kappa}=\frac12\langle p_1,p_1\rangle
+\frac12\langle p_2,p_2\rangle -(1+2\kappa)\langle p_1,e_2\rangle
\langle p_2,e_1\rangle-\tr(e_1e_2^T\Xi). \label{pendulum2}
\end{equation}
are given by
\begin{eqnarray}
&& \dot e_1=p_1-(1+2\kappa)\langle e_1,p_2\rangle  e_2, \nonumber\\
&& \dot e_2=p_2-(1+2\kappa)\langle e_2,p_1\rangle e_1, \label{1}\\
&& \dot p_1=+\Xi e_2 + (1+2\kappa)\langle p_1,e_2\rangle p_2+\lambda_{11} e_1+ \lambda_{12} e_2, \nonumber\\
&& \dot p_2=-\Xi e_1 + (1+2\kappa)\langle p_2,e_1\rangle
p_1+\lambda_{21} e_1+ \lambda_{22} e_2, \nonumber
\end{eqnarray}
where the Lagrange multipliers are:
\begin{equation}
\begin{pmatrix}
\lambda_{11} & \lambda_{12}\\
\lambda_{21}& \lambda_{22}
\end{pmatrix}=\begin{pmatrix}
-\langle p_1,p_1\rangle -\langle e_1,\Xi e_2\rangle  & -\langle p_1,p_2\rangle \\
-\langle p_1,p_2\rangle  & -\langle p_2,p_2\rangle -\langle
e_1,\Xi e_2\rangle
\end{pmatrix}.
\label{2}
\end{equation}

The integrability  of the system \eqref{1}, \eqref{2} essentially
follows from the construction given in \cite{BJ4}, which we
briefly recall below.

\subsection{Pendulum systems on adjoint orbits of compact Lie groups}

Let $G$ be a compact semisimple Lie group. The pendulum system on
the adjoint orbit $\mathcal O(a)$ is defined by the Hamiltonian
function of the form
\begin{equation}\label{pendulum-orbit}
H(x,p)=\frac12\langle\Phi_0(x,p),\Phi_0(x,p)\rangle-\langle
x,b\rangle,
\end{equation}
where $x=\Ad_g(a)\in\mathcal O(a)$, $p\in T^*_x\mathcal O(a)$,
$\Phi_0: T^*\mathcal O(a)\to\g$ is the momentum mapping of the
natural Hamiltonian $G$-action, and $b\in\g$ is a fixed element of
the Lie algebra. Here we identified $\g$ with its dual space
$\g^*$ by means of the standard invariant scalar product
$\langle\cdot,\cdot\rangle$, the Killing form multiplied by $-1$.

Let $ \mathfrak g^\C=\mathfrak g \otimes \mathbb{C}. $ Then
$\mathfrak g^\C$ is a semisimple complex Lie algebra. Denote by
$\mathfrak g_0$ the real semisimple Lie algebra obtained from
$\mathfrak g^\C$: $ \mathfrak g_0=\mathfrak g \oplus  \sqrt{-1}
\mathfrak g. $ Then $\dim \mathfrak g_0=\dim_\R \mathfrak
g^\C=2\dim_\C \mathfrak g^\C=2\dim \mathfrak g$ and $ \rank_\C
\mathfrak g^\C=\rank \mathfrak g=r$, $\rank \mathfrak g_0 = 2\rank
\mathfrak g=2r.
$

Let $p_1,\dots,p_r$ be the set of basic homogeneous invariant
polynomials on $\mathfrak g$ considered as complex invariant
polynomials on $\mathfrak g^\C$. Then their real and imaginary
parts form a set of basic  polynomial invariants on $\mathfrak
g_0$.

The pair $(\mathfrak g_0,\mathfrak g)$ is symmetric ($[\mathfrak
g,\sqrt{-1}\mathfrak g]\subset \sqrt{-1}\mathfrak g$,
$[\sqrt{-1}\mathfrak g,\sqrt{-1}\mathfrak g]\subset\mathfrak g$),
and, as in the previous section, one can consider the contraction
of $\mathfrak g_0$: the real Lie algebra $\mathfrak g \oplus_{\ad}
 \sqrt{-1}\mathfrak g$,  the semidirect product where the second
term is considered as a commutative subalgebra.

Let us identify $(\mathfrak g \oplus_{\ad} \sqrt{-1} \mathfrak
g)^*$ with $\mathfrak g \oplus_{\ad} \sqrt{-1} \mathfrak g$ by
means of nondegenerate scalar product $
(\xi_1+\sqrt{-1}\eta_1,\xi_2+\sqrt{-1}\eta_2)=\langle
\xi_1,\xi_2\rangle - \langle \eta_1,\eta_2 \rangle$, which is
proportional to the Killing form of $\mathfrak g_0$. Then the
Lie-Poisson brackets on $(\mathfrak g \oplus_{\ad} \sqrt{-1}
\mathfrak g)^*$ becomes
\begin{eqnarray*}
\{f,g\}_{0}(\xi+i\eta) = \langle \xi, [\nabla_\xi f,\nabla_\xi
g]\rangle + \langle \eta, [\nabla_\xi f,\nabla_\eta g] +
[\nabla_\eta f,\nabla_\xi g]\rangle.
\end{eqnarray*}

Again, from a general construction of compatible Poisson
structures related to a symmetric pair decomposition of Lie
algebras  we conclude that
\begin{equation}
\mathcal A= \{\mathfrak{Re}(p_j(\lambda
\xi+\sqrt{-1}(\eta+\lambda^2 b)),\, \mathfrak{Im}(p_j(\lambda
\xi+\sqrt{-1}(\eta+\lambda^2 b)),
 \, \lambda\in\R,\, j=1,\dots,r\}
\label{B}
\end{equation}
is a complete commutative set on $(\mathfrak g \oplus_{\ad} \sqrt{-1}
\mathfrak g)^*$ (see \cite{BJ4}).

The mapping $ \Upsilon_0: T^*\mathcal O(a) \to (\mathfrak g
\oplus_{\ad} \sqrt{-1} \mathfrak g)^*$ given by
\begin{equation}
\Upsilon_0(x,p)=\Phi_0(x,p)+\sqrt{-1} x \label{theta}
\end{equation}
is a symplectomorphism between $T^*\mathcal O(a)$ and the
coadjoint orbit of the element $a$ in $(\mathfrak g \oplus_{\ad}
\sqrt{-1} \mathfrak g)^*$ endowed with the canonical
Kirillov-Konstant symplectic form. Moreover, for an arbitrary $a$
(regular or singular), the set $\mathcal A$ is complete on the
image $\Upsilon_0(T^*\mathcal O(a))$, for a generic $b\in\g$ (see
\cite{BJ4}).

\begin{thm}[\cite{BJ4}]\label{pomoc}
Let $a\in\g$ be an arbitrary element of the Lie algebra $\g$ of a
compact semi-simple Lie group $G$. For a generic $b\in\g$, the
pendulum system on the adjoint orbit $\mathcal O(a)$ defined by
the Hamiltonian \eqref{pendulum-orbit} is Liouville integrable by
means of the integrals $\Upsilon_0^*\mathcal A$.
\end{thm}

\subsection{Integrability}

The flow (\ref{1}), (\ref{2}) is right $SO(2)$-invariant and has
the momentum integral $\Psi_{12}$.


\begin{thm} \label{LA_pair2}
1). The equations \eqref{1}, \eqref{2} imply the following matrix equation with
a spectral parameter $\lambda$
\begin{eqnarray}
&&\frac{d}{dt}
L(\lambda)=[A(\lambda),L(\lambda)],\label{lax1}\\
&& L(\lambda)=\lambda \Phi + \sqrt{-1}\, ( e_1\wedge e_2+\lambda^2
\Xi), \quad A(\lambda)=\Phi+\sqrt{-1}\, \lambda \Xi.\nonumber
\end{eqnarray}
2). For a generic $\Xi\in so(n)$, the
pendulum system \eqref{1}, \eqref{2} is completely integrable.
\end{thm}

\noindent{\it Proof.} 1). From (\ref{1}) it follows
$$
\frac{d}{dt}(e_1\wedge e_2)=[\Phi,e_1 \wedge e_2],\quad
\frac{d}{dt}\Phi=[e_1\wedge e_2,\Xi],
$$
which implies the Lax representation (\ref{lax1}).

\

2). The mapping
$$
\Upsilon: \quad (e_1,e_2,p_1,p_2) \longmapsto \Phi + \sqrt{-1}\,
e_1 \wedge e_2=p_1 \wedge e_1+ p_2 \wedge e_2 + \sqrt{-1}\, e_1
\wedge e_2
$$
is the Poisson mapping between $(T^*V(n,2),\omega)$ and the dual
space of the semi-direct product $so(n) \oplus_{\ad} \sqrt{-1}
so(n)$.

Now let $\mathfrak F$ be the union of the real and imaginary parts of the
coefficients of the spectral curve
$$
\det(L(\lambda)-\mu\mathbf{I}_{n})=0.
$$
They are the pull-backs, via the mapping $\Upsilon$, of the
commuting polynomials \eqref{B} on $(so(n) \oplus_{\ad} \sqrt{-1}
so(n))^*$ with $b=\Xi$. Besides, since $L(\lambda)$ is
$SO(2)$-invariant, the integrals $\mathfrak F$ are
$SO(2)$-invariant as well. Therefore, the collection of functions
given by $\mathfrak F$ and $\Psi_{12}$ is commutative.

It remains to estimate the number of independent integrals. Let
$\hat\pi$ be a natural extension of \eqref{pr} to the
$SO(2)$-projection $\Psi_{12}^{-1}(0)\to T^*G^+(n,2)$. Then
\begin{equation}\label{pomoc2}
\Upsilon\vert_{\Psi_{12}^{-1}(0)}=\Upsilon_0\circ\hat\pi,
\end{equation}
where $\Upsilon_0: T^*G^+(n,2)\to (so(n) \oplus_{\ad} \sqrt{-1}
so(n))^*$ is given by \eqref{theta}.

Next, applying Theorem \ref{pomoc} we get that, for a generic $\Xi$,
there are $2n-4=\dim G^+(n,2)$ polynomials $f_1,\dots,f_{2n-4}$ in
$\mathcal A$, such that the functions
$$
\Upsilon_0\circ f_1,\dots, \Upsilon_0\circ f_{2n-4}\,: \quad
T^*G^+(n,2)\to\R
$$
are independent. In addition, from \eqref{pomoc2}, we conclude that the functions
\begin{equation}\label{pomoc3}
F_1=\Upsilon\circ f_1,\dots, F_{2n-4}=\Upsilon\circ
f_{2n-4}\,:\quad T^*V(n,2)\to\R
\end{equation}
are independent at the invariant set $\{\Psi_{12}=0\}\subset
T^*V(n,2)$. Since all the functions are analytic, taking
\eqref{pomoc3} and $\Psi_{12}$ we obtain $2n-3=\dim V(n,2)$
commuting independent integrals, which implies the commutative integrability. \hfill$\Box$

\begin{remark}{\rm One can generalize the pendulum system on $V(n,2)$ by adding the
closed 2-form $\eta\, de_1 \wedge de_2=\eta \sum_{i=1}^n de_1^i
\wedge de_2^i$ to the standard symplectic structure, describing a
magnetic force field. Then Theorem \ref{LA_pair2} still holds,
with the momentum mapping $\Phi$ replaced by the magnetic momentum
mapping $ \Phi_{\eta}=\Phi+\eta \, e_1 \wedge e_2=p_1 \wedge e_1+
p_2 \wedge e_2 + \eta\,  e_1 \wedge e_2. $ }\end{remark}

\section{The fourth degree potential}

\subsection{Invariant relations}
Consider the system on $V(n,r)$ with the following fourth degree
potential and the kinetic energy $T_\kappa$:
\begin{equation}
H(X,P)=T_\kappa+\frac12\tr (X^T B^2 X)-\frac 12\tr(X^TBXX^TBX),
\label{4s}\end{equation} where $B=\diag(b_1,b_2,\dots,b_n)$. For
the case $r=1$, $b_i=\sqrt{2} a_i$, $i=1,\dots,n$ we get the
Hamiltonian (\ref{4sp}) with $\nu=1$.

As follows from the Dirac equations,
the  flow of the Hamiltonian function \eqref{4s} is given by
\begin{equation}
\begin{aligned}
&\dot X=P-(1+2\kappa)XP^TX , \\
&\dot
P=(1+2\kappa)PX^TP-B^2X+2BXX^TBX+X\Lambda,\end{aligned}\label{S_XP}
\end{equation}
where $\Lambda=X^TBX-2X^TBXX^TBX-P^TP$ is the matrix Lagrange multiplier.

The Hamiltonian (\ref{4s}) is $SO(r)$-invariant. Therefore the
momentum mapping (\ref{momentum_right_map}) is an integral of the
motion.

\begin{lem}
The derivation of the $SO(n)$-momentum mapping
\eqref{momentum_map} along the flow \eqref{S_XP} reads
$$
\frac{d}{dt}\Phi=\frac12[B,(\mathbf{I_n}-2XX^T)B(\mathbf{I_n}-2XX^T)].
$$
Further, on the invariant subvariety
\begin{equation}
\Psi^{-1}(0): \qquad \{(X,P)\, | \,X^TP=P^TX=0 \}
\label{invariant}
\end{equation}
the additional relation holds:
$$\frac{d}{dt}\left((\mathbf{I}_n-2XX^T)B(\mathbf{I}_n-2XX^T)\right)=2[\Phi,(\mathbf{I}_n-2XX^T)B(\mathbf{I}_n-2XX^T)].$$
\end{lem}

The formulas of the lemma lead to

\begin{thm} \label{LA_pair3}
On the invariant submanifold \eqref{invariant}, the equations
\eqref{S_XP} imply the matrix equation with a spectral parameter
$\lambda$
\begin{eqnarray}
&&\frac{d}{dt} L(\lambda)=[L(\lambda),A(\lambda)], \label{LAS}\\
&& L(\lambda)=(\mathbf{I_n}-2XX^T)B(\mathbf{I_n}-2XX^T)+2\lambda
\Phi+\lambda^2B, \quad A(\lambda)=-2\Phi-\lambda B.\nonumber
\end{eqnarray}
\end{thm}

Therefore, the functions
\begin{equation}
\{\, L(\lambda)^k\, \vert\,  \lambda\in \R, \, k=1,\dots,n\, \}
\label{C}
\end{equation}
are {\it particular integrals} of the system (\ref{S_XP}) on the
invariant manifold (\ref{invariant}).

\subsection{Integrability on the Grassmannian variety}

The system \eqref{S_XP}, the Lax representation (\ref{LAS}), as
well as the functions (\ref{C}), are $SO(r)$-invariant and are
well defined on the reduced Poisson space $(T^*V(n,r))/SO(r)$.
After $SO(r)$-symplectic reduction of the system to
$\Psi^{-1}(0)/SO(r)\thickapprox T^* G^+(n,r)$ we get a natural
mechanical system with a kinetic energy given by the normal
metric. 

\begin{thm}
The Hamiltonian flow \eqref{S_XP} reduced to the cotangent bundle
of the the oriented Grassmannian $G^+(n,r)$ is completely
integrable.
\end{thm}

\noindent{\it Proof.} Let $\mathcal{G}(n,r)\subset SO(n)$ be the
Cartan model of the {\it nonoriented} Grassmannian variety
$G(n,r)$. Consider the {\it generalized Neumann system} on
$\mathcal G(n,r)$ with the kinetic energy $K$ induced from the
bi-invariant metric on $SO(n)$ and the potential function
$W(g)=\langle \Ad_g(B),B\rangle$, $g\in \mathcal G(n,r)$ (see
\cite{S1, S2}).\footnote{Note that one should distinguish the
notion of generalized Neumann systems on Cartan models of
symmetric spaces with the co-adjoint representations of the
Neumann systems on the sphere $S^{n-1}$ \cite{Ra,Ju} and on
Stiefel manifolds $V(n,r)$, $r>1$ \cite{FeJo3}.} The complete
integrability of this system follows from general Saksida's
construction \cite{S1, S2}, which we adopt for the case of the
Cartan model for the Grassmannian. The latter has the following
description. Let $\mathbf
J_{r,n-r}=\diag(-\mathbf{I}_r,\mathbf{I}_{n-r})$. The mapping
$$
\theta: G^+(n,r) \to \mathcal{G}(n,r), \quad
\theta(\pi(X))=(\mathbf{I}_n-2XX^T)\mathbf J_{r,n-r}
$$
is the double-covering map from the oriented Grasmannian variety
to the Cartan model of the non-oriented one (e.g., see \cite{Fo,
Jo2}). Let $\tilde\theta$ be the natural extension of $\theta$ to
the mapping of corresponding cotangent bundles.

The pull-back of the potential
\begin{eqnarray*}
V=(\pi\circ\theta)^* W &=&\langle\Ad_{(\mathbf{I}_n-2XX^T)\mathbf J_{r,n-r}} B,B\rangle \\
&=& \langle (\mathbf{I}_n-2XX^T)\mathbf J_{r,n-r}B((\mathbf{I}_n-2XX^T)\mathbf J_{r,n-r})^T ,B\rangle \\
 &=&-\frac12\tr\left((\mathbf{I}_n-2XX^T)\mathbf
J_{r,n-r}B\mathbf J_{r,n-r}(\mathbf{I}_n-2XX^T)B\right) \\&=& 2\tr
(X^T B^2 X)-2\tr(X^TBXX^TBX)-\frac12\tr B^2,
\end{eqnarray*}
up to addition of a constant, is proportional to that given in
\eqref{4s}, while the pull-back $\tilde\theta^*K$ of the kinetic
energy of Saksida's model is the kinetic energy given by a normal
metric on $G^+(n,r)$. Besides, the integrals (\ref{C}) (considered
on $T^*G^+(n,r)$) are the pull-backs of the integrals of the
generalized Neumann system on $T^*\mathcal{G}(n,r)$ given in
\cite{S1}. Thus, since the mapping $\tilde\theta^*$ preserves the
commutativity and independency of functions (see Proposition 3 in
\cite{S2}), the integrals (\ref{C}) imply the complete integrability
of the reduced flow on $T^*G^+(n,r)$.\hfill$\Box$

\begin{remark}{\rm It is proved that the reduced system on the Poisson manifold
$(T^*V(n,r))/SO(r)$ is integrable on the single symplectic leaf.
 The situation is similar to what happens in several classical systems, in particular in the
{\it Goryachev-Chaplygin case} of a heavy rigid body
motion, when an additional integral exist and the system is
solvable only for the zero value of the $SO(2)$-momentum mapping (see, e.g., \cite{Go}).}
\end{remark}

\subsection*{Acknowledgments}
The research of B. J. was supported by the Serbian Ministry of
Science, Project {174020, Geometry and Topology of Manifolds,
Classical Mechanics and Integrable Dynamical Systems}. The
research of Yu. F. was supported by the MICINN-FEDER grants
MTM2009-06973, MTM2009-12670, and CUR-DIUE grant 2009SGR859.


\small
\sc

\

Yuri N. Fedorov

Department de Matem\`atica I

Universitat Politecnica de Catalunya

Barcelona, E-08028 Spain

{\rm e-mail: Yuri.Fedorov@upc.es}

\

Bo\v zidar Jovanovi\' c

Mathematical Institute SANU

Serbian Academy of Sciences and Arts

Kneza Mihaila 36, 11000, Belgrad, Serbia

{\rm e-mail: bozaj@mi.sanu.ac.rs}


\begin{thebibliography}{99}

\bibitem{BCS} {Bloch A M , Crouch P E, Sanyal A K} 2006  A variational problem on Stiefel manifolds,
Nonlinearity, {\bf 19}, 2247-2276.

\bibitem{Bo} Bolsinov A V 1991 {Compatible Poisson brackets on Lie
algebras and the completeness of families of functions in involution},
Izv. Acad. Nauk SSSR, Ser. matem. {\bf 55}, No.1, 68-92
(Russian); English translation: Math. USSR-Izv., {\bf 38},  No.1, 69-90  (1992).

\bibitem{Bo2}
Bolsinov A V 1991 Commutative families of functions related to
consistent Poisson brackets. Acta Appl. Math. {\bf 24}, no. 3,
253–-274.

\bibitem{BJ} Bolsinov A V, Jovanovi\' c B 2001
{Integrable geodesic flows on homogeneous spaces}. {Matem.
Sbornik} {\bf 192} no. 7, 21-40 (Russian); English translation:
{Sb. Mat.} {\bf 192}, no. 7--8, 951--968 (2001).


\bibitem{BJ4} Bolsinov A V, Jovanovi\' c B 2008
Magnetic Flows on Homogeneous Spaces, Comm. Math. Helv., Comment.
Math. Helv. {\bf 83} (2008), no. 3, 679--700,
arXiv:math-ph/0609005.

\bibitem{Dirac} Dirac P A 1950
On generalized Hamiltonian dynamics. Can. J. Math.
{\bf 2}, no.2, 129--148.

\bibitem{FeJo3} Fedorov Yu N, Jovanovi\' c B  Geodesic Flows and Neumann Systems on Stiefel Varieties.
Geometry and Integrabilty, Mathematische Zeitschrift,
Mathematische Zeitschrift  DOI: 10.1007/s00209-010-0818-y,
arXiv:1011.1835.

\bibitem{Fo}
Fomenko A T 1983 {\it Differential Geometry and Topology.
Supplementary Chapters}, Moscow University, Moscow, 1983, 217 p.
(Russian)

\bibitem{Go} Golubev V V 1953
\emph{Lectures on integration of the equations of motion of a
rigid body about a fixed point}, Moskva, Gostenhizdat, (in
Russian); English translation: Transl. Philadelphia, PA: Coronet
Books, 1953.

\bibitem{GS} Guillemin V and Sternberg S 1984
{\it Symplectic techniques in physics.} Cambrige University press.

\bibitem{Je} {Jensen G} 1973 Einstein metrics on principal
fiber bundles, J. Diff. Geom. {\bf 8}, 599-614.

\bibitem{Jo2} Jovanovi\' c B 2007
On the Cartan Model of the Canonical Vector Bundles over
Grassmannians, Sib. Mat. Zh. Vol. 48, No. 4, 772--777 (Russian);
English translation. Siberian Mathematical Journal, Vol. 48
(2007), No. 4., 616--620, arXiv:math/0602132

\bibitem{Ju} Jurdjevic V 2011 Integrable Hamiltonian Systems on
Symmetric Spaces: Jacobi, Kepler and Moser, arXiv:1103.2818



\bibitem{Moser} Moser J 1980
Geometry of quadric and spectral theory. In: Chern Symposium 1979,
Berlin--Heidelberg--New York, 147--188.

\bibitem{Pa} Panasyuk A 2009
Bi-Hamiltonian structures with symmetries, Lie pencils and
integrable systems.  J. Phys. A  {\bf 42},  no. 16, 165205.

\bibitem{Ra} {Ratiu T} 1981  The C. Neumann problem as a completely integrable
system on an adjoint orbit, Trans. Amer. Math. Soc. {\bf 264},
no.2, 321-329.

\bibitem{R} Reyman A G 1980 Integrable
Hamiltonian systems connected with graded Lie algebras,
Zap. Nauchn. Semin. LOMI AN SSSR {\bf 95}, 3-54 (Russian);
English translation: J. Sov. Math. {\bf 19}, 1507-1545, (1982).

\bibitem{RS}
Reyman A G and Semonov-Tian-Shanski M A 1994
Group theoretical methods in the theory of finite dimensional integrable systems.
In. {\it Dynamical systems VII} (Eds.: V. I. Arnold, S. P. Novikov),  Springer.

\bibitem{S1} Saksida P 1999 Nahm's equations and generalizations of the Neumann system,
Proc. Lond. Math. Soc. {\bf 78}, 701-720.

\bibitem{S2} Saksida P 2001 Integrable anharmonic oscilators on spheres and hyperbolic spaces.
Nonlinearity {\bf 14}, 977-994.

\bibitem{TF} Trofimov V V, Fomenko A T 1995
{\it Algebra and geometry of integrable Hamiltonian differential
equations.} Moskva, Faktorial, (Russian).

\bibitem{Wo} Wojciechowski S 1985 Integrable one-partical potentials
related to the Neumann system and the Jacobi problem of geodesic
motion on an ellipsoid, Phys. Lett. A {\bf 107} 107-111.

\end{thebibliography}
\end{document}